\documentclass[superscriptaddress,prd,aps,onecolumn,floats,amsfonts,amssymb,10pt]{revtex4}

\usepackage{mathrsfs,graphicx}

\begin{document}

\title{Sensitivity of quintessence perturbations to initial conditions}

\author{Jingsong Liu}
\affiliation{Dept.~of Physics, Stanford University, Stanford, CA 94305-4060}

\date{\today}

\begin{abstract}

Gauge invariant quintessence perturbations in the quintessence and cold
dark matter ($QCDM$) model are investigated. 
For three cases of constant equation-of-state (EOS) parameter, linear
scalar field potential, and supergravity scalar field potential, their
perturbation evolutions have a similar dependence on EOS parameter and
scale, but they have different sensitivity to the initial conditions due
to the different shapes of the quintessence potential. They have a minor
effect on primary CMB anisotropies, but change the secondary CMB effect in
different ways. The first case is insensitive to initial quintessence
perturbations and only modifies the Integrated Sachs-Wolfe (ISW) effect
within a factor of 2. The other two cases are sensitive to initial
conditions at large scales and could affect the secondary CMB anisotropies
drastically depending on how smooth the initial perturbations are. This
makes it possible for future cosmological probes to provide constraints on
quintessence properties. 

\end{abstract}

\maketitle


\section{Introduction}

Recent observations of an accelerating universe \cite{Riess1998,
Perlmutter1999, Riess2001} imply the existence of dark energy
characterized by a negative pressure to density ratio, also known as the
EOS parameter $w$. There have been discussions of the cosmological
constant, the dark energy evolving according to a specific EOS parameter
\cite{Silveira1997, Chiba1997}, and the dark energy consisting of a
dynamical cosmic scalar field, the quintessence \cite{Ratra1988,
Ferreira1997}. In contrast to the cosmological constant, the quintessence
component in $QCDM$ model has kinematic behavior and can develop
perturbations. It has important features of time-evoluting EOS parameter
and scale-dependent effective sound speed \cite{Caldwell1998a,
Caldwell1998b}. One significant issue is the sensitivity to initial
perturbation conditions, which attracts much attention. 

For some $QCDM$ models, the quintessence potentials can be approximated
with constant EOS parameter, and the perturbations evolutions are
insensitive to the initial quintessence perturbations, with minor effect
on observations \cite{Dave2002, DaveThesis, Weller2003}. For exponential
potential $V(Q)=\hat{V} e^{-(c/M)Q}$ \cite{Wetterich1985, Wetterich1988,
Ratra1988}, although quintessence perturbations may stay non-zero at some
time for different background attractor solutions, they die out in
matter-dominant time for physically reasonable models \cite{Tassilo2001},
which shows the insensitivity again. For the tracking quintessence models
\cite{Brax2000}, large-scale non-adiabatic perturbations can grow or stay
constant before entering tracker regime, but have to get suppressed
afterwards \cite{MalquartiLiddle2002}. However, the first-order matrix
formulation developed latter \cite{Bartolo2003} applies to general
quintessence potential cases on all linear scales and indicates the
possibility of non-vanishing entropy perturbations. Recently some other
$QCDM$ models with special quintessence potentials, such as linear $V(Q)$
\cite{Dimopoulos2003} and supergravity $V(Q)$ \cite{Kallosh2002}, lead to
special background evolutions. So it will be interesting to see the
perturbations behavior and their sensitivity to initial conditions in
these models, as well as their effects on cosmic microwave background
(CMB) \cite{Bennett2003} and baryon mass power spectrum
\cite{Scranton2003} in the universe. 

In this paper, we describe the universe by a simple $QCDM$ model with the
conformal-Newtonian metric in section 2. The cosmological perturbations
are solved for constant $w$ cases in section 3, for linear-$V(Q)$ cases
in section 4, and for supergravity-$V(Q)$ cases in section 5. The results
are analyzed by the matrix formulation and their effects on cosmological
observations are discussed in section 6.

\section{Quintessence and Cold Dark Matter (QCDM) universe}

Here we consider the uniform matter and quintessence field background with
small perturbations in a perturbed flat FRW metric. Focusing on scalar
perturbations, we can use the flat conformal-Newtonian metric for the
gauge invariant approach \cite{Bardeen1980, Mukhanov1992}: 
\begin{eqnarray}
\label{ConformalNewtonianMetric}
ds^2 = a^2(\eta) \left[ -(1+2\Phi(\eta,\vec{x})) d\eta^2 +
  (1-2\Phi(\eta,\vec{x})) \delta_{ij} dx^i dx^j \right]
\end{eqnarray}
where $\eta$ is the conformal time and $|\Phi(\eta,\vec{x})| \ll 1$ is the
gravitational potential. We define 
$^{\prime} \equiv \frac{\partial}{\partial \eta}$, 
$\mathcal{H} \equiv \frac{a^{\prime}}{a}$
as the ``Hubble parameter'' in the conformal-Newtonian version,
$\eta_0$ as today's conformal time, and 
$\bar{a} \equiv \frac{a(\eta)}{a(\eta_0)}$ as the normalized scale factor.
For the quintessence scalar field, we assume the standard Lagrangian
$\mathcal{L}_Q = -\frac{1}{2} Q_{,\mu} Q^{,\mu} - V(Q)$
and decompose the quintessence field as uniform background with small
spatial perturbations: 
$Q(\eta, \vec{x}) \equiv Q_0(\eta) + \delta Q(\eta, \vec{x})$,
$|\delta Q| \ll |Q_0|$, $\langle \delta Q(\eta, \vec{x}) \rangle = 0$,
where $\langle\;\rangle$ means global spatial average.
We assume that the quintessence part and matter part only interact
with each other gravitationally. So the wave equation of the quintessence
field 
$-Q^{;\mu}_{\; ;\mu} + \partial V / \partial Q = 0$
gives its background and perturbative components:
\begin{eqnarray}
& & Q_0^{\prime\prime} + 2\mathcal{H} Q_0^{\prime} + a^2 V_{,Q}(Q_0) = 0
\label{QEquation.0} \; , \\
& & \delta Q^{\prime\prime} + 2\mathcal{H} \delta Q^{\prime} 
- \nabla^2 \delta Q + a^2 V_{,QQ} \delta Q 
- 4 Q_0^{\prime} \Phi^{\prime} + 2 a^2 V_{,Q} \Phi = 0
\label{QEquation.1} \; . 
\end{eqnarray}
The energy-momentum tensor of this quintessence field is
$T_Q^{\mu\nu} = Q^{,\mu} Q^{,\nu} 
+ (-\frac{1}{2} Q_{,\sigma} Q^{,\sigma} - V) \; g^{\mu\nu}$.

For the matter part, we assume its energy-momentum tensor has the form of the
perfect fluid: 
$T^{\mu \nu} \equiv (\rho_m + p_m) U^{\mu} U^{\nu} + p_m \, g^{\mu \nu}$,
where 
$U^{\mu} \equiv d x^{\mu} / d\tau = [ (1-\Phi)/a, V^i/a ]$
is the 4-velocity of the fluid,
$V^i \equiv d x^i / d\eta = a\, d x^i / dt \ll 1$
is its 3-velocity, 
$\rho_m(\eta, \vec{x}) \equiv 
\langle\rho_m\rangle(\eta) [1 + \delta_m(\eta, \vec{x})]$,
and $|\delta_m(\eta, \vec{x})| \ll 1$.
If we neglect all pressure effect 
($|p_m(\eta, \vec{x})| \ll \rho_m(\eta, \vec{x})$, 
$|\nabla\delta p_m| \ll \rho_m |\nabla\Phi|$) 
on scales which are not too small ($>10$ Mpc), the
energy-momentum-conservation law gives background and perturbation 
components after some manipulation: 
\begin{eqnarray}
& & \langle\rho_m\rangle = \frac{\langle\rho_m\rangle (\eta_0)}{\bar{a}^3}
\label{MatterEMC.1} \; , \\
& & \delta_m^{\prime \prime} + \mathcal{H} \delta_m^{\prime} 
- 3\Phi^{\prime \prime} - 3\mathcal{H} \Phi^{\prime} - \nabla^2 \Phi = 0
\label{MatterEMC.4} \; .
\end{eqnarray}
The matter perturbation $\delta_m$ can be formally obtained in terms of $a$
and $\Phi(k, \eta)$ in $k$-space, where $k$ is the comoving wave-vector
($-\nabla_{\vec{x}}^2 = k^2$).

To complete this system, we consider the 0th-order 0-0 component and the
1st-order $i$=$j$ component of the Einstein equations: 
\begin{eqnarray}
& & \frac{3\mathcal{H}^2}{a^2} = 
8\pi G \left[ \frac{\langle \rho_m \rangle (\eta_0)}{\bar{a}^3} 
+ \frac{Q_0^{\prime \, 2}}{2a^2} + V(Q_0) \right]
\label{EE.0.1} \; , \\
& & \Phi^{\prime\prime} + 3\mathcal{H} \Phi^{\prime} + 
(2\mathcal{H}^{\prime} + \mathcal{H}^2) \Phi = 
4\pi G a^2 \left( -\frac{Q_0^{\prime \, 2}}{a^2} \Phi
+ \frac{Q_0^{\prime} \delta Q^{\prime}}{a^2} - V_{,Q} \delta Q \right) 
\label{EE.1.3} \; .
\end{eqnarray}
Given the initial conditions, $V(Q)$, and other parameters,
eq.(\ref{QEquation.0}, \ref{EE.0.1}) give the background $Q_0(\eta)$ and
$a(\eta)$, and eq.(\ref{QEquation.1}, \ref{EE.1.3}) give the perturbations
$\delta Q$ and $\Phi$. Due to the convenient conformal-Newtonian metric,
the matter perturbation $\delta_m$ drops out and can be found later via
eq.(\ref{MatterEMC.4}).

\section{Constant $w$ case}

\begin{figure}[ht]
\includegraphics[scale=0.75]{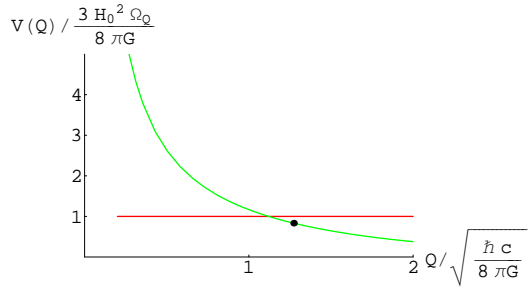}
\caption{Scalar field potential $V(Q)$ for constant-$w$ cases. The curved
line is for $w=-2/3$. The scalar field rolls from top-left corner down to
the today's value indicated by the dot. The horizontal line is for $w=-1$
($\Lambda CDM$).} 
\label{ConstantW_V_Figure}
\end{figure}

Here we follow the EOS parameterization in \cite{Dave2002} to study the
general properties of the quintessence field instead of some specific
quintessence field potential $V(Q)$.
\begin{eqnarray}
w(\eta) \equiv \frac{\langle p_Q \rangle}{\langle \rho_Q \rangle} = 
\frac{\frac{Q_0^{\prime \, 2}}{2a^2} - V(Q_0)}{\frac{Q_0^{\prime \, 2}}{2a^2} + V(Q_0)}
\label{EOS} \; .
\end{eqnarray}
Then we have
\begin{eqnarray}
\frac{Q_0^{\prime \, 2}}{2a^2} = \frac{1+w}{1-w} V(Q_0) 
= \frac{1+w}{2} \langle \rho_Q \rangle.
\label{EOS2}
\end{eqnarray}
By manipulating eq.(\ref{QEquation.0}, \ref{EOS2}), we can express
$V_{,Q}$ and $V_{,QQ}$ in terms of the EOS parameter $w(\eta)$:
\begin{eqnarray}
& & a^2 V_{,Q} = -\frac{Q_0^{\prime}}{2} \left[ 3(1-w)\mathcal{H} 
+ \frac{w^{\prime}}{1+w} \right]
\label{V'} \; , \\
& & a^2 V_{,QQ} = -\frac{3}{2} (1-w) 
\left[ \frac{a^{\prime\prime}}{a} - \mathcal{H}^2 (\frac{7}{2}+\frac{3}{2}w) \right]
+ \frac{1}{1+w} \left[ \frac{w^{\prime 2}}{4(1+w)} -
\frac{w^{\prime\prime}}{2} + w^{\prime} \mathcal{H} (3w+2) \right]
\label{V''} \; . 
\end{eqnarray}
With eq.(\ref{V'}, \ref{V''}) and definitions 
\begin{eqnarray}
& & Q_0^{\prime} \equiv a(\eta_0) \sqrt{(1+w)
\langle \rho_Q \rangle(\eta_0)} \; \psi^{\prime},
\;\;\;\; \psi^{\prime}(\eta_0) = 1,
\label{Q_0Redefinition} \\
& & \delta Q \equiv a(\eta_0) \sqrt{(1+w)
\langle \rho_Q \rangle(\eta_0)} \; \delta \psi,
\label{DeltaQRedefinition}
\end{eqnarray}
we can put the quintessence wave equations eq.(\ref{QEquation.0},
\ref{QEquation.1}) in a more convenient form:
\begin{eqnarray}
& & \psi^{\prime\prime} + \frac{1+3w}{2} \mathcal{H} \, \psi^{\prime} = 0,
\label{QEquationConstantW.0} \\
& & \delta \psi^{\prime\prime} 
+ (2\mathcal{H} + \frac{w^{\prime}}{1+w}) \delta \psi^{\prime}
+ \left\{ -\nabla^2 - \frac{3}{2} (1-w) \left[ \frac{a^{\prime\prime}}{a} 
- \mathcal{H}^2 (\frac{7}{2} + \frac{3w}{2}) \right] + 3\mathcal{H} w^{\prime}
\right\} \delta \psi 
\nonumber \\
& & \;\;\;\;\;\;\;\; - 4\psi^{\prime} \Phi^{\prime} 
- \left[ 3\mathcal{H} (1-w) + \frac{w^{\prime}}{1+w} \right] \psi^{\prime}
\Phi = 0.
\label{QEquationConstantW.1}
\end{eqnarray}
In this formulation, any function $V(Q)$ is equivalently described by the
corresponding EOS parameter $w(\eta)$, which can be much more easily specified
and compared with other dark energy models. One simple example would be
$w=constant$, which has the exponential-like potential $V(Q)$ shown in
Figure~\ref{ConstantW_V_Figure}. 

In this special case, eq.(\ref{EOS2}, \ref{Q_0Redefinition},
\ref{QEquationConstantW.0}) give the solution for the quintessence field
background: 
\begin{eqnarray}
& & \psi^{\prime} = \bar{a}^{-(1+3w)/2} \; , 
\label{ConstantEOSpsiSolution} \\
& & \langle \rho_Q \rangle = 
\frac{\langle \rho_Q(\eta_0) \rangle}{\bar{a}^{3(1+w)}} \; .
\label{ConstantEOSvarepsilonSolution} 
\end{eqnarray}
Then the previous closed system of $a$, $Q_0^{\prime}$, $\Phi$ and $\delta Q$
can be reduced to a system of $a$, $\Phi$ and $\delta \psi$ in $k$-space:
\begin{eqnarray}
& & \frac{\mathcal{H}^2}{\bar{a}^2} = 
(H_0 a_0)^2 \left[ \frac{\Omega_m}{\bar{a}^3} 
+ \frac{\Omega_Q}{\bar{a}^{3(1+w)}} \right] ,
\label{ClosedSystem.1} \\
& & \Phi^{\prime\prime} + 3\mathcal{H} \Phi^{\prime} + 
(2\mathcal{H}^{\prime} + \mathcal{H}^2) \Phi = 
\frac{3}{2} (H_0 a_0)^2 (1+w) \Omega_Q \left[ -\frac{\Phi}{\bar{a}^{1+3w}} 
+ \frac{\delta \psi^{\prime}}{\bar{a}^{(1+3w)/2}} 
+ \frac{3}{2} \mathcal{H} (1-w) \frac{\delta \psi}{\bar{a}^{(1+3w)/2}} \right], 
\label{ClosedSystem.2} \\
& & \delta \psi^{\prime\prime} + 2\mathcal{H} \delta \psi^{\prime}
+ \left\{ k^2 - \frac{3}{2} (1-w) \left[ \frac{a^{\prime\prime}}{a} 
- \mathcal{H}^2 (\frac{7}{2} + \frac{3w}{2}) \right] \right\} \delta \psi 
- 4 \frac{\Phi^{\prime}}{\bar{a}^{(1+3w)2}} 
- 3\mathcal{H} (1-w) \frac{\Phi}{\bar{a}^{(1+3w)/2}} = 0,
\label{ClosedSystem.3}
\end{eqnarray}
where $\Omega_m$ and $\Omega_Q$ are the density parameters defined as usual. 
If $w=-1$, the quintessence perturbations vanish in eq.(\ref{ClosedSystem.2}) 
and only its background contributes. This special case thus reduces to the
$\Lambda CDM$ model. At early times, this model can be well approximated
by the flat $CDM$ model (Einstein-DeSitter space). Then eq.(\ref{ClosedSystem.1})
gives $\bar{a}(\eta) \propto \eta^2$, $\mathcal{H} \simeq \frac{2}{\eta}$, and
eq.(\ref{ClosedSystem.2}) gives $\Phi$ independent of time for the 
growing modes of matter perturbations $\delta_m \propto \bar{a}(\eta)$, as
usual. At later times, the quintessence part begins to become important
and the solution of $\bar{a}(\eta)$ from eq.(\ref{ClosedSystem.1}) will
deviate from $\eta^2$ more and more as time goes on. So the coefficient
$2\mathcal{H}^{\prime} + \mathcal{H}^2$ in eq.(\ref{ClosedSystem.2}) is not
zero any more and makes $\Phi$ decay. This is what causes the ISW effect
in low-$l$ multipoles of the CMB power spectrum. 

Recent cosmological observation of the accelerating universe requires
$w(\eta_0)<-1/3$. The $w<-1$ cases would correspond to negative kinetic
term in the standard quintessence Lagrangian or other more complicated
models, which we won't discuss here. So from now on we just consider
$-1 \leq w<-1/3$ in all our $QCDM$ models. The initial conditions are $a(0) \cong 0$,
$|\Phi(0)| \sim 10^{-5}$ and $\Phi^{\prime}(0)=0$ for modes outside
horizon, based on inflation theory \cite{LindePPIC}. For the same reason,
$|\delta\psi(0)| \sim |\Phi(0)|$.  
Because in eq.(\ref{ClosedSystem.3}) the restoring term ($V_{,QQ} \propto
\mathcal{H}^2$) is bigger than the damping term ($\propto \mathcal{H}$)
which is bigger than the driving term ($\propto \mathcal{H} \, 
\bar{a}^{-(1+3w)/2}$) at early times, $\delta\psi$ with any initial
value which is not too big ($|\delta\psi(0) / \Phi(0)| \leq 10^4$) will
oscillate and get damped quickly without affecting the $\Phi$ evolution
before the last-scattering epoch. So the primary effects on the CMB won't get
modified much and are insensitive to the initial condition on the
quintessence perturbations. This is one important result within many
$QCDM$ models \cite{Dave2002, DaveThesis}. Here we can see that it is mainly due to
the huge $V_{,QQ}$, i.e., the huge quintessence mass at early times. 
For now we just take the smooth initial condition $\delta\psi(0) =
\delta\psi^{\prime}(0) = 0$ for convenience. Let us look at the behavior
of this system at various spatial scales. 

At small scales ($k \gg 1$), by eq.(\ref{ClosedSystem.3}) the growth of
$\delta\psi$ is strongly suppressed and stays negligible compared with
$\Phi$. The effective sound speed of the quintessence perturbations approaches 1
($c_s^2 \equiv \delta p_{Q} / \delta\rho_{Q} \simeq 1$). Then we can drop
the $\delta \psi$ terms in the above system and the evolution of $\Phi$
only comes from background effect without a dependence on $k$ in this region:
\begin{eqnarray*}
& & \Phi^{\prime\prime} + 3\mathcal{H} \Phi^{\prime} + 
(2\mathcal{H}^{\prime} + \mathcal{H}^2) \Phi = 
- \frac{3}{2} (H_0 a_0)^2 (1+w) \frac{\Omega_Q}{\bar{a}^{1+3w}} \Phi \; .
\label{BigKClosedSystem.2}
\end{eqnarray*}
This system behaves like the $\Lambda CDM$ model with a different EOS parameter
$w$ and the additional $r.h.s$ term. If $w$ is bigger the dark energy
takes over dominance earlier and $\bar{a}$ deviates from $\eta^2$ earlier
too, which makes $\Phi$ decay more. In addition, the negative $r.h.s$
brings down $\Phi$ more. So compared with the $\Lambda CDM$ model,
small-scale quintessence perturbations in the $QCDM$ model do not grow and
only the background effect contributes, which drags down the 
gravitational potential more and corresponds to a bigger ISW effect.

At scales comparable to or larger than horizon size ($k \leq 1$), the
quintessence perturbations $\delta \psi$ can grow to the order of the 
gravitational potential $\Phi$ by eq.(\ref{ClosedSystem.3}) with smaller 
effective sound speed ($c_s^2<1$). Then it will back-react on the $\Phi$
evolution via eq.(\ref{ClosedSystem.2}). Because of the positive
coefficient, the quintessence perturbations tend to raise the
gravitationally potential $\Phi$ to make it decay less. This is different
from the $k \gg 1$ case, and agrees with the result of a phenomenological
approach \cite{Bean2003}. Thus in this $QCDM$ model the evolution of
perturbations and gravitational potential depends on scale. 

\begin{figure}[ht]
\includegraphics[scale=0.75]{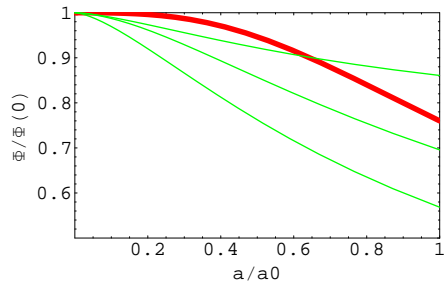}
\caption{Time evolution of the gravitational potential $\Phi(\eta)$ for the
constant-$w$ case. The three thin lines correspond to $k=\{0.01,5,500\}$ 
from top to bottom for the $w=-0.5$ case. The thick line is the
scale-independent $w=-1$ case ($\Lambda CDM$). } 
\label{ConstantW_Phi_Figure}
\end{figure}

With definition on horizon scale $a_0 \equiv \frac{2}{H_0}$, numerical
calculation gives $\Phi(\eta)$ and $\delta\psi(\eta)$ once we specify the
parameters $k$ and $w$. Corresponding to the assumptions we made before for
this linearized system, we only consider $k<800$. For $w=-0.5$ and
$k=\{0.01, 5, 500\}$, the time evolution $\Phi(\eta)$ is shown compared
with the $w=-1$ case in Figure~\ref{ConstantW_Phi_Figure}. 

\section{Linear $V(Q)$ case}

\begin{figure}[ht]
\includegraphics[scale=0.75]{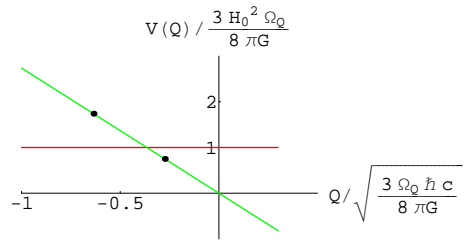}
\caption{Linear scalar field potential $V(Q)$ for the $w(\eta_0)=-0.5$ case
shown as the slope. The two dots are the starting and today's positions of
the field variable from left to right. The horizontal line is the $w=-1$ case
($\Lambda CDM$).} 
\label{LinearV_V_Figure}
\end{figure}

Linear scalar field potential was looked by Andrei \cite{HawkingGravitation} in
inflation theory long ago. Recently it is reviewed by Dimopoulos and Thomas 
\cite{Dimopoulos2003} as a quintessence field within a quantum
mechanically stable action. Due to the large Z-moduli of the wave function
factor, the higher order potential derivatives $d^n V / dQ^n \; (n>1)$ can
be neglected and only the linear term remains. After shifting the origin
and considering the symmetry of $Q \leftrightarrow -Q$, the typical $V(Q)$
can be written as: 
\begin{eqnarray}
V(Q) = - V_1 \; Q, \;\;\;\;\;\;\;\;V_1 = constant > 0.
\label{LinearV}
\end{eqnarray}
This linear potential may eventually lead to negative total energy density
and result in a rapid crunch. Let us consider the closed system defined in
$\S$2 to study the behavior of both background and perturbations for this
model in detail. In order of magnitude, Eq.(\ref{QEquation.0}) gives 
$Q_0^{\prime} \sim a_0^2 \eta_0 V_1 \sim a_0 V_1 / H_0$
and eq.(\ref{EOS2}) gives 
$Q_0^{\prime \, 2} / 2a^2 \sim |V_1 Q| \sim \langle \rho_Q \rangle
\sim 3 H_0^2 \Omega_Q / 8\pi G$
for $1+w \sim 1$. Combining these two, we will have
$V_1 \sim H_0^2 \sqrt{3\Omega_Q / 8\pi G} \sim H_0^2 m_p \; (\hbar=c=1)$
and $Q \sim \sqrt{3\Omega_Q / 8\pi G} \sim m_p$.
So this simple analysis sets the range of magnitude for the quintessence
field quantities. Now we can redefine the quintessence field for later
convenience: 
\begin{eqnarray}
V_1 = H_0^2 \sqrt{\frac{3\Omega_Q}{8\pi G}} \; \tilde{V}_1, \;\;\;\;
Q = \sqrt{\frac{3\Omega_Q}{8\pi G}} \; \tilde{Q}, \;\;\;\;
\tilde{V}_1 \sim \tilde{Q} \sim 1, \;\; \tilde{V}_1 = constant > 0.
\label{QRedefined}
\end{eqnarray}
The closed system of equations can then be rewritten as:
\begin{eqnarray}
& & \tilde{Q}_0^{\prime\prime} + 2\mathcal{H} \tilde{Q}_0^{\prime} 
- H_0^2 a_0^2 \bar{a}^2 \tilde{V}_1 = 0 \; ,
\label{QEquationLinearV.0} \\
& & \frac{\bar{a}^{\prime}}{\bar{a}^2} = H_0 a_0 
\sqrt{\frac{\Omega_m}{\bar{a}^3} + \Omega_Q \left( 
\frac{\tilde{Q}_0^{\prime \, 2}}{2H_0^2 a_0^2 \bar{a}^2} 
- \tilde{V}_1 \tilde{Q}_0 \right)} \; ,
\label{EERedefinedLinearV.0.1} \\
& & \delta \tilde{Q}^{\prime\prime} + 2\mathcal{H} \delta \tilde{Q}^{\prime} 
+ k^2 \delta \tilde{Q} - 4 \tilde{Q}_0^{\prime} \Phi^{\prime} 
- 2 H_0^2 a_0^2 \bar{a}^2 \tilde{V}_1 \Phi = 0 \; ,
\label{QEquationLinearV.1} \\
& & \Phi^{\prime\prime} + 3\mathcal{H} \Phi^{\prime} + 
(2\mathcal{H}^{\prime} + \mathcal{H}^2) \Phi = 
\frac{3}{2} \Omega_Q \left( -\tilde{Q}_0^{\prime \, 2} \Phi
+ \tilde{Q}_0^{\prime} \delta \tilde{Q}^{\prime} 
+ H_0^2 a_0^2 \bar{a}^2 \tilde{V}_1 \delta \tilde{Q} \right) \; ,
\label{EERedefinedLinearV.1.3}
\end{eqnarray}
where the perturbation equations are in $k$-space. For the background,
we take the initial condition as $\bar{a}(0) = 0$ and
$\tilde{Q}_0^{\prime}(0) = 0$ based on inflation theory. The
$\tilde{Q}_0(0)$ value is chosen by requiring that today's quintessence
energy density fits the observation, 
$\rho_Q(\eta_0) = 3H_0^2 \Omega_Q / 8\pi G$,
or equivalently, 
$\tilde{Q}_0^{\prime \, 2} / 2H_0^2 a_0^2 - \tilde{V}_1 \tilde{Q}_0(\eta_0) = 1$.
Then this describes an initially static field rolling down a potential
slope. Its effective EOS parameter evolves from -1 to some value
$-1<w(\eta_0)<0$ today. Different slope values $\tilde{V}_1$ correspond to
different $\tilde{Q}_0(0)$ and EOS parameter $w(\eta)$, but the same energy
density $\rho_Q(\eta_0)$ today. The correspondence is listed in
Table~\ref{V-w_Table}. The detail has been worked out in the ``doomsday''
model \cite{Kallosh2003}. For the $w(\eta_0)=-0.5$ case, the potential $V(Q)$
is shown in Figure~\ref{LinearV_V_Figure}. 

\begin{table}[ht] 
\caption{\label{V-w_Table}Quintessence potential slope and starting position
corresponding to $w(\eta_0)$} 
\begin{tabular}{ccc} \hline \hline
$\tilde{V}_1$	& $\tilde{Q}_0(0)$	& $w(\eta_0)$	\\  \hline
2.738		& -0.635		& -0.5		\\
2.424		& -0.657		& -0.6		\\
2.077		& -0.696		& -0.7		\\
1.677		& -0.774		& -0.8		\\
1.172		& -0.980		& -0.9		\\
\hline \hline
\end{tabular}
\end{table}

For the perturbations, again we take the initial condition as $\delta
\tilde{Q}^{\prime}(0) = \Phi^{\prime}(0) = 0$ and $|\Phi(0)| \sim 10^{-5}$,
based on inflation theory. Eq.(\ref{QEquationLinearV.1}) describes the
oscillation of an intitially static quintessence perturbation field driven
by the gravitational potential. At small scales ($k \gg 1$) the quintessence
perturbations stay small. So the $\Phi$ evolution only depends on the
background. The behavior of the perturbations is not sensitive to $k$ or
$\delta\tilde{Q}(0)$, as in the constant-$w$ case. However, at scales 
comparable or bigger than the horizon size ($k \leq 1$), due to
the zero quintessence mass ($V_{,QQ}=0$) in the restoring term, the 
quintessence perturbation will stay near its initial value with a 
shift driven by the gravitational potential. Then the whole system becomes
sensitive to the initial condition of quintessence perturbations, which is
quite different from the constant-$w$ case. For the $w(\eta_0) = -0.5$
special case, the dependence on $k$ and $\delta \tilde{Q}(0)$ is indicated
in Table~\ref{LinearV_Initial_Value_Dependence_Table}.

\begin{table}[ht] 
\caption{\label{LinearV_Initial_Value_Dependence_Table} Dependence on
initial perturbations in the linear-$V(Q)$ case, for $w(\eta_0)=-0.5$} 
\begin{tabular}{|c|c|c|c|c|c|c|}  \hline \hline
& \multicolumn{2}{|c|}{$k=0.1$} & \multicolumn{2}{|c|}{$k=1$} & \multicolumn{2}{|c|}{$k=10$} \\ 
\hline
{$\delta\tilde{Q}(0) / \Phi(0)$} & 
{$\delta\tilde{Q}(\eta_0) / \Phi(0)$} & {$\Phi(\eta_0) / \Phi(0)$} & 
{$\delta\tilde{Q}(\eta_0) / \Phi(0)$} & {$\Phi(\eta_0) / \Phi(0)$} & 
{$\delta\tilde{Q}(\eta_0) / \Phi(0)$} & {$\Phi(\eta_0) / \Phi(0)$} \\ 
\hline
 10  &  12.077  &  3.862  &  9.411  &  3.262  &  0.147  &  0.741  \\
  1  &   1.752  &  1.116  &  1.469  &  1.054  &  0.108  &  0.730  \\
  0  &   0.605  &  0.811  &  0.586  &  0.808  &  0.104  &  0.729  \\
 -1  &  -0.542  &  0.506  & -0.296  &  0.563  &  0.099  &  0.728  \\
-10  & -10.867  & -2.240  & -8.240  & -1.646  &  0.060  &  0.717  \\
\hline \hline
\end{tabular}
\end{table}

\begin{figure}[ht]
\includegraphics[scale=0.75]{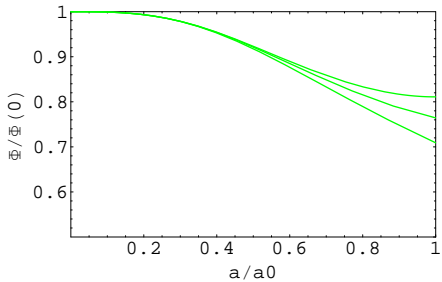}
\caption{Time evolution of the gravitational potential $\Phi(\eta)$ for the
linear-$V(Q)$ case with $w(\eta_0)=-0.5$ and smooth initial condition. The
three curves correspond to $k=\{0.01, 5, 500\}$ from top to bottom.}
\label{LinearV_Phi_Figure}
\end{figure}

If we take the smooth initial condition $\delta\tilde{Q}(0)=0$, then in terms
of different $k$ the perturbation evolution result is shown in
Figure~\ref{LinearV_Phi_Figure}. Again we see that the scale dependence of  
perturbations evolutions is similar to the constant-$w$ case in the last section. 
The quintessence perturbations stay suppressed, corresponding to more decay
of the gravitational potential at small scales, but grow to the order of the
gravitational potential accounting for less decay of the gravitational
potential, with the transition scale around $k=5$. Besides, the decay of
the gravitational potential is smaller than the constant-$w$ case because its
quintessence EOS parameter stays close to -1 at early times and only rises
to $w(\eta_0)$ recently, so that the dark energy emerges later. 

\section{Supergravity V(Q) case}

\begin{figure}[ht]
\includegraphics[scale=0.75]{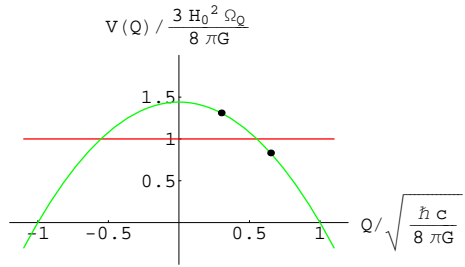}
\caption{Scalar field potential $V(Q)$ from supergravity for
$\tilde{Q}_0(0)=0.3$ shown as the curve. The two dots are the starting and
today's positions of the field variable, from left to right. The horizontal
line is the $w=-1$ case ($\Lambda CDM$).} 
\label{SugraV_V_Figure}
\end{figure}

The scalar field potential can also come from extended gauged supergravity,
which has a close relation to M/string theory and extra dimensions. Masses of
ultra-light scalars in these models \cite{Kallosh2002} are quantized by
the Hubble constant: $m^2=n^2H^2$. If the de-Sitter solution corresponds to a
minimum of the effective potential, the universe eventually becomes 
de-Sitter space. If the de-Sitter solution corresponds to a maximum or a
saddle point, which is the case in all known models based on $N=8$
supergravity, the flat universe eventually stops accelerating and
collapses to a singularity. For example, all known potentials of the
gauged $N=8$ supergravity have the universal feature $n^2=\{-6,4,12\}$. 
Along the tachyon direction $n^2=-6$, the scalar potential can be written as:
\begin{eqnarray}
& & V(Q) = \frac{3H^2_0 \Omega_Q \tilde{V}_0}{8\pi G} (1-\tilde{Q}^2),
\label{SugraV}  \\
& & \tilde{V}_0 = constant > 0, 
\;\;\;\; \tilde{Q} \equiv \frac{Q}{m_p}, 
\;\;\;\; m_p \equiv \sqrt{\frac{\hbar c}{8\pi G}} \nonumber
\end{eqnarray}
where $\tilde{V}_0$ and $\tilde{Q}$ are dimensionless and in the order of 1. 
Again we apply the closed system defined in $\S$2 to this model. The
background and perturbations equations are:
\begin{eqnarray}
& & \tilde{Q}_0^{\prime\prime} + 2\mathcal{H} \tilde{Q}_0^{\prime} 
- 6 H_0^2 a_0^2 \Omega_Q \tilde{V}_0 \bar{a}^2 \tilde{Q}_0 = 0 \; ,
\label{QEquationSugraV.0} \\
& & \frac{\bar{a}^{\prime}}{\bar{a}^2} = 
\sqrt{\frac{H_0^2 a_0^2 \Omega_m}{\bar{a}^3} 
      + \frac{\tilde{Q}_0^{\prime \, 2}}{6\bar{a}^2} 
      + H_0^2 a_0^2 \Omega_Q \tilde{V}_0 \tilde{Q}_0 (1-\tilde{Q}_0^2)} \; ,
\label{EERedefinedSugra.0.1} \\
& & \delta \tilde{Q}^{\prime\prime} + 2\mathcal{H} \delta \tilde{Q}^{\prime} 
+ (k^2 - 6 H_0^2 a_0^2 \bar{a}^2 \Omega_Q \tilde{V}_0) \delta \tilde{Q} 
- 4 \tilde{Q}_0^{\prime} \Phi^{\prime} 
- 12 H_0^2 a_0^2 \bar{a}^2 \Omega_Q \tilde{V}_0 \tilde{Q}_0 \Phi = 0 \; ,
\label{QEquationSugraV.1} \\
& & \Phi^{\prime\prime} + 3\mathcal{H} \Phi^{\prime} + 
(2\mathcal{H}^{\prime} + \mathcal{H}^2) \Phi = 
- \frac{1}{2} \tilde{Q}_0^{\prime \, 2} \Phi
+ \frac{1}{2} \tilde{Q}_0^{\prime} \delta \tilde{Q}^{\prime} 
+ 3 H_0^2 a_0^2 \bar{a}^2 \Omega_Q \tilde{V}_0 \tilde{Q}_0 \delta \tilde{Q} \; .
\label{EERedefinedSugra.1.3}
\end{eqnarray}
For the background, we again take the initial condition as $\bar{a}(0) = 0$ and
$\tilde{Q}_0^{\prime}(0) = 0$. We choose $\tilde{Q}_0(0) = \{0, 0.2, 0.3,
0.32 \}$ and adjust $\tilde{V}_0$ so that 
today's quintessence energy density fits the observation, 
$\rho_Q(\eta_0) = 3H_0^2 \Omega_Q / 8\pi G$,
or equivalently, 
$\tilde{Q}_0^{\prime \, 2} / 6H_0^2 a_0^2 \Omega_Q 
+ \tilde{V}_0 [1-\tilde{Q}_0^2(\eta_0)] = 1$.
Then this describes an initially static field rolling down a curved potential
slope. Its effective EOS parameter evolves from -1 to some value
$-1<w(\eta_0)<0$ today. Different initial positions $\tilde{Q}_0(0)$ correspond to
different $\tilde{V}_0$ and EOS parameter $w(\eta)$, but the same energy
density $\rho_Q(\eta_0)$ today. The correspondence is listed in
Table~\ref{Q(0)-w_Table}. For the $\tilde{Q}_0(0)=0.3$ case, the potential $V(Q)$ is
shown in Figure~\ref{SugraV_V_Figure}. 

\begin{table}[ht] 
\caption{\label{Q(0)-w_Table}Quintessence initial position corresponding
to $\tilde{V}_0$ and $w(\eta_0)$} 
\begin{tabular}{ccc} \hline \hline
$\tilde{Q}_0(0)$	&  $\tilde{V}_0$	&  $w(\eta_0)$	\\  \hline
0			&  1			&  -1		\\
0.2			&  1.124		&  -0.92	\\
0.3			&  1.441		&  -0.67	\\
0.32			&  1.645		&  -0.47	\\
\hline \hline
\end{tabular}
\end{table}

The perturbations are analyzed as before. We take the initial condition as
$\delta \tilde{Q}^{\prime}(0) = \Phi^{\prime}(0) = 0$ and 
$|\Phi(0)| \sim 10^{-5}$. Eq.(\ref{QEquationSugraV.1}) describes the
oscillation of an initially static quintessence perturbation field driven
by the gravitational potential. At small scales ($k \gg 1$) any initial
quintessence perturbations get damped to be small, and the $\Phi$
evolution only depends on the background. This system is not sensitive to 
$\delta \tilde{Q}(0)$, as in the linear-$V(Q)$ case. At scales comparable
to or bigger than the horizon size ($k \leq 1$), due to the tachyon quintessence
mass ($V_{,QQ} < 0$) in the restoring term, a quintessence perturbation
will stay near its initial value and even grow, with a shift driven by the
gravitational potential. Then this system is more sensitive to the initial
quintessence perturbations than the linear-$V(Q)$ case. For the
$\tilde{Q}_0(0) = 0.3$ case, the dependence on $k$ and $\delta
\tilde{Q}(0)$ is indicated in Table~\ref{SugraV_Initial_Value_Dependence_Table}. 

\begin{table}[ht] 
\caption{\label{SugraV_Initial_Value_Dependence_Table} Dependence on
initial perturbations in Supergravity-$V(Q)$ case, for $\tilde{Q}_0(0)=0.3$} 
\begin{tabular}{|c|c|c|c|c|c|c|}  \hline \hline
& \multicolumn{2}{|c|}{$k=0.1$} & \multicolumn{2}{|c|}{$k=1$} & \multicolumn{2}{|c|}{$k=10$} \\ 
\hline
{$\delta\tilde{Q}(0) / \Phi(0)$} & 
{$\delta\tilde{Q}(\eta_0) / \Phi(0)$} & {$\Phi(\eta_0) / \Phi(0)$} & 
{$\delta\tilde{Q}(\eta_0) / \Phi(0)$} & {$\Phi(\eta_0) / \Phi(0)$} & 
{$\delta\tilde{Q}(\eta_0) / \Phi(0)$} & {$\Phi(\eta_0) / \Phi(0)$} \\ 
\hline
 10  &  23.412  &  3.016  &  18.640  &  2.589  &  0.207  &  0.753  \\
  1  &   3.018  &  1.011  &   2.521  &  0.967  &  0.159  &  0.746  \\
  0  &   0.752  &  0.788  &   0.731  &  0.787  &  0.153  &  0.746  \\
 -1  &  -1.514  &  0.565  &  -1.060  &  0.606  &  0.148  &  0.745  \\
-10  & -21.908  & -1.440  & -17.179  & -1.016  &  0.100  &  0.739  \\
\hline \hline
\end{tabular}
\end{table}

\begin{figure}[ht]
\includegraphics[scale=0.75]{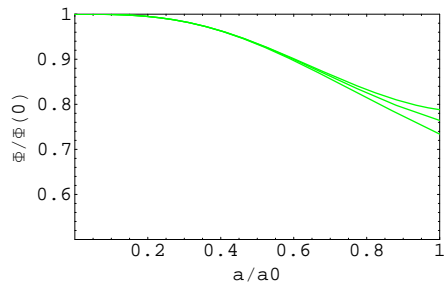}
\caption{Time evolution of the gravitational potential $\Phi(\eta)$ for the
supergravity-$V(Q)$ case, with $\tilde{Q}_0(0)=0.3$ and smooth initial
condition. The three curves correspond to $k=\{0.01, 5, 500\}$ from top to bottom.} 
\label{SugraV_Phi_Figure}
\end{figure}

This potential is unbounded from below and the theory is unstable. Since the
potential can remain positive for $|\Phi| < 1$ for a time longer than the
present age of the universe before it finally collapses, this model is
sufficient to describe the background evolution of the universe. However,
the tachyon mass makes the quintessence instability develop and affect
the gravitational potential, and thus the system becomes sensitive to the
initial conditions at large scales ($k<10$). This puts more constraints on
the initial quintessence conditions and this supergravity model itself. 
For the smooth initial condition $\delta\tilde{Q}(0)=0$, this model has a 
similar scale-dependent perturbation evolution as the linear-$V(Q)$ model
in last section. For $\tilde{Q}_0(0)=0.3$ and $k=\{0.01, 5, 500\}$, the
time evolution $\Phi(\eta)$ is shown in Figure~\ref{SugraV_Phi_Figure}. 

\section{Discussion}
In another viewpoint, we can apply the first-order matrix formulation in
\cite{Bartolo2003} to analyze the different quintessence perturbation
behaviors in above $QCDM$ models. At large scales ($k \ll 1$), the relative 
entropy perturbation $S$ and intrinsic entropy perturbation $\Gamma$
\cite{Wands2000, Malik2003, Kodama1984} have coupled evolution equation 
$$
\frac{\partial}{\partial \ln(a/a_0)}
\left( \begin{array}{c} S \\ \Gamma \end{array} \right)
= 3 \left( \begin{array}{cc}
w_Q - w_f + \gamma_f \Omega_f (w_f - c_{sQ}^2) / \gamma 
& \gamma_f \Omega_f (1 - c_{sQ}^2) / \gamma \\
- \gamma /2 & w_Q - \gamma /2  
\end{array} \right)
\times \left( \begin{array}{c} S \\ \Gamma \end{array} \right)
$$
where $_f$ means the perfect fluid component, $\gamma_i \equiv 1+w_i$, and
$c_{sQ}^2 \equiv \dot{p} / \dot{\rho}$. For most of the time in our
$QCDM$ models, $CDM$ dominates and have $w_f \simeq 0$, $\Omega_f \simeq
\gamma_f \simeq \gamma \simeq 1$, and $-1<w_Q<-1/3$. Then the eigenvalues
are $n_{\pm} = \frac{3}{2} \left[ 2w_Q - c_{sQ}^2 - \frac{1}{2} \pm
\sqrt{(c_{sQ}^2 + \frac{1}{2})^2 - 2} \right]$. For the constant $w$ case,
$w_Q = c_{sQ}^2$ leads to $Re(n_{\pm}) = \frac{3}{2}(w_Q - 1/2) < 0$,
corresponding to decaying entropy perturbations. While for the linear
$V(Q)$ case, $c_{sQ}^2 \simeq -2$ leads to $n_{\pm} = (3w_Q+3,3w_Q+\frac{3}{2})$,
corresponding to at least one growing entropy perturbation mode. This fits
our previous analysis well. 

In all $QCDM$ models that we studied above, the quintessence perturbations
have the common characteristic feature of scale dependence. At small
scales the perturbations do not grow and the gravitational potential
decays more only due to background effects; at large scales the 
perturbations grow and sustain the gravitational potential. On the other 
hand, because of the different quintessence potentials, the three $QCDM$
models have different sensitivity to the initial conditions on quintessence
perturbations. The constant-$w$ model has a large quintessence mass at
early times and damps perturbations fast so that it is insensitive to
initial conditions at all scales. In this model the quintessence has no
effect before the last-scattering time. In contrast, the linear-$V(Q)$
model and supergravity-$V(Q)$ model have zero and tachyon quintessence
mass respectively, and their perturbations remain near their initial value
($\delta\tilde{Q}(0) / \Phi(0) \leq 10^4$) at large scales. Since the EOS
parameter $w$ in these two models approaches -1 at early times, their dark
energy dominates only very recently, like the $\Lambda CDM$ model. So the
total back-reaction of dark energy perturbations (background times
relative perturbations) on other perturbations remains negligible before
the last-scattering time, when the universe evolves like $a \propto
\eta^2$, $Q_0 \simeq Q_0(0)$, $Q_0^{\prime} \simeq 0$, $\delta Q \simeq
\delta Q(0)$, and $\Phi \simeq \Phi(0)$. So they do not change the primary
CMB anisotropies either. But later when the dark energy background become
dominant, this back-reaction of dark energy perturbations will affect
other perturbations drastically, depending on how smooth the initial
perturbations are. 

\begin{figure}[ht]
\includegraphics[scale=0.75]{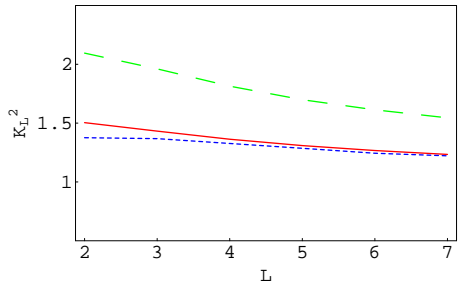}
\caption{The low-$l$ CMB power spectrum due to the Sachs-Wolfe effect in
various $QCDM$ models with the smooth initial condition. The solid line is
for the $\Lambda CDM$ model, the dash line for the constant-$w$ $QCDM$
model with $w=-0.6$, and the dot line for the linear-$V(Q)$ $QCDM$ model
with $w(\eta_0)=-0.5$.} 
\label{ISW_Figure}
\end{figure}

For smooth initial conditions in all models the dark energy has no
primary CMB effect except that its background contributes to the distance
to the last-scattering surface. But later on both the dark energy
background and perturbations will contribute to the change of the
gravitational potential, which may lead to a considerable secondary CMB 
effect such as the ISW effect \cite{SW1967}. For the scale-invariant
primordial spectrum $P_{\Phi} \equiv |\Phi(0,k)|^2 = A^2/k^3$, the
large-scale CMB power spectrum due to the SW effect is: 
\begin{eqnarray}
& & C_l^{SW} = \frac{A^2}{36\pi^2 l(l+1)} K_l^2 \; , \nonumber \\
& & K_l^2 = 2l(l+1) \int_0^{\infty} \frac{dk}{k} \left[ 
j_l(k\eta_0) + 6\int_{\eta_{LS}}^{\eta_0} d\eta f^{\prime}(\eta,k) j_l(k(\eta_0-\eta))
\right]^2 \; ,
\label{ISW_K}
\end{eqnarray}
where $j_l(z)$ is the spherical Bessel function and $f(\eta, k)$ is the
evolution factor of the gravitational potential defined as $\Phi(\eta, k) 
\equiv \Phi(0, k) f(\eta, k)$. The integral over the $j_l^2(k\eta_0)$ term
corresponds to the primary SW effect on the last-scattering surface, and
the integral over the remaining terms corresponds to the ISW effect. The
flat $CDM$ model gives $f^{\prime}=0$ and only the primary SW effect, with
$K_l^2=1$. The flat $\Lambda CDM$ model has no dark energy perturbations
and gives a scale-independent evolution factor $f(\eta)$ which leads to a
positive ISW effect in the low-$l$ CMB power spectrum \cite{Kofman1985}, 
rising at the lower-$l$ ($l=2,3$) end. For the flat $QCDM$ model, the
bigger EOS parameter $w$ $(\geq -1)$ brings out the dark energy earlier
and decreases the gravitational potential more, corresponding to a bigger
$f^{\prime}$ value for all scales and affecting the magnitude of the CMB
power spectrum. The quintessence perturbations grow differently at 
different scales and give a $k$-dependent $|f^{\prime}(\eta,k)|$, which is
bigger at small scales than at large scales with the transition around the
horizon size. This means that the lower-$l$ ($l=2,3$) multipoles will get
suppressed relative to the higher-$l$ ones such that the whole low-$l$
spectrum will get flattened. These two effects can be seen by numerical
calculation for the  standard $\Lambda CDM$ model, constant-$w$ $QCDM$
model with $w=-0.5$, and linear-$V(Q)$ $QCDM$ model with $w(\eta_0)=-0.5$
in Figure~\ref{ISW_Figure}. This is done with eq.(\ref{ISW_K})
approximately for illustrative purpose. Exact results should refer to more
accurate numerical work like the CMB-FAST code. For non-smooth initial
conditions, the perturbation evolution can be seen in
Table~\ref{LinearV_Initial_Value_Dependence_Table} and 
Table~\ref{SugraV_Initial_Value_Dependence_Table}, and corresponds to very
unusual cosmological scenarios which we will not discuss further.

The scale dependence and sensitivity to initial perturbations can also be
tested in other ways. The CMB results give the primordial $\Phi_k$ spectrum, 
while the gravitational lensing and other large-scale observations yield
$\Phi_k$ information at low redshift. The ratio of these two, the
evolution of gravitational potential, can be directly compared with the
theoretical plot on different scales shown above for various $QCDM$ models. 
Furthermore, since the matter density perturbation is related to the
gravitational potential by eq.(\ref{MatterEMC.4}), the evolution of
gravitational potential can also be inferred from the measurement of the
visible matter density spectrum (by SDSS, etc.), and checked with $QCDM$ predictions.

By considering the dark energy perturbations in these $QCDM$ models, we
see that they all have a scale dependence which is different from the
$\Lambda CMD$ model, and they have different sensitivity to initial
conditions. These properties make the dark energy perturbations an
effective way to test these models. More precise cosmological observations
in the future will give better constraints on these $QCDM$ models.

\section{Acknowledgements}

This work is supported by NASA grant NAS 8-39225 to Gravity Probe B. The
author is grateful to F.Everitt, R.Wagoner, R.Adler, A.Silbergleit, and
the Gravity Probe B theory group for their valuable remarks.



\end{document}